# A New View on the Application of Gold Nanoparticles in Cancer Therapy

A. M. Dadabaev[*], Yu. Kh.-M. Shidakov, V. M. Lelevkin, A. A. Sorokin, K. A. Moldosanov[*]

(Kyrgyz-Russian Slavic University named after B. N. Yeltsin, Bishkek, Kyrgyzstan)

[*]Corresponding authors: azim.dadabaev.science@gmail.com, kamil.moldosanov@mail.ru

In biomedical research and the practice of cancer therapy, gold nanoparticles have been used to visualize malignant tumors, as the heated bodies for hyperthermia of cancer cells, as drug carriers to deliver drugs to a cancer cell, but, to the best of our knowledge, they have not yet been used consciously as the sources of terahertz (THz) radiation delivered to a cancer cell that contributes to the inhibition of cell activity. It is predicted here that gold nanoparticles ≤ 8 *nm* in size are sources of spontaneous THz radiation, and the possibility of their application in oncology is due to the known effects of THz radiation on the cells of living organisms. There are indications that nanoparticles with a size comparable to the width of the major groove of the DNA molecule will be the most effective. Another effect that has not yet been taken into account in biomedical studies using gold nanoparticles is that of local electric fields due to the contact potential difference above edges and vertices of gold nanoclusters. The prerequisites and possibilities for searching for the manifestations of these two effects when gold nanoparticles are introduced into living cells of organisms are considered.

-----

## 1. Введение. О новом взгляде на применение наночастиц золота в онкологии

Хотя следы применения коллоидного золота в лечебных практиках уходят в глубь веков [1] – оно упоминается ещё в трактатах китайских, арабских и индийских учёных V и IV веков до нашей эры – потенциал его применения, очевидно, ещё не исчерпан. Успехи в понимании эффектов, связанных с применением золотых наночастиц (ЗНЧ) в медицине, достигнутые за последние тридцать лет, освещены, например, в обзорах [1-10]. Здесь же мы укажем на два физических явления, которые ещё не учитывались при использовании ЗНЧ в биомедицинских исследованиях (вероятно, вследствие малого перекрывания сфер исследовательской деятельности двух международных биомедицинских сообществ: «*nanoparticle community*» и «*terahertz community*»). Между тем, учёт этих явлений мог бы пролить новый свет на процессы, возможные в живой клетке при внесении в неё ЗНЧ. Для удобства рассмотрения напомним сначала о трёх эффектах, происходящих в клетке живого организма, облучаемой ТГц фотонами:

*(а)* локальное разъединение цепочек у двухцепочечной молекулы ДНК в результате локального разрыва водородных связей между парами азотистых оснований аденин–тимин и гуанин–цитозин, из-за чего между цепочками образуются локальные «пузыри» (*bubbles*) [11]; это согласуется с данными [12] о поглощении ТГц излучения азотистыми основаниями (Фиг. 1), а также с кривой общего поглощения молекулы ДНК [13] (Фиг. 2);

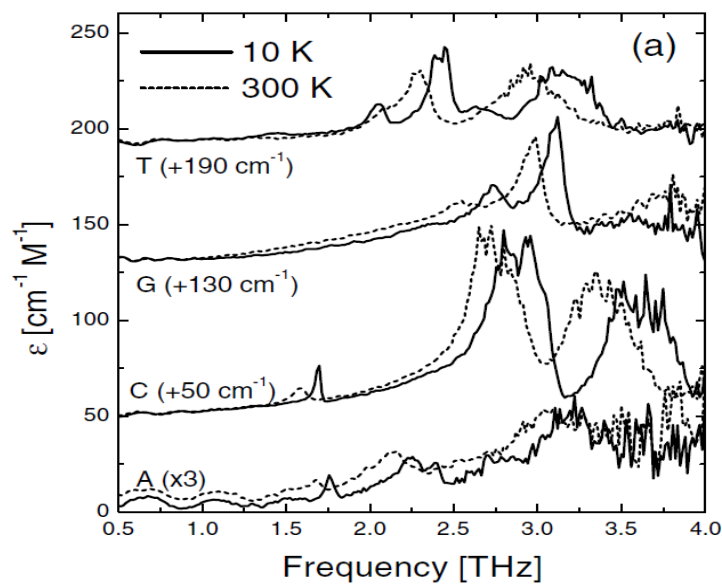

Фиг. 1. Коэффициенты поглощения молекул, соединяющих спирали ДНК:
аденина (A), гуанина (G), цитозина (C) и тимина (T) при температурах 10 *K* и 300 *K* [12].

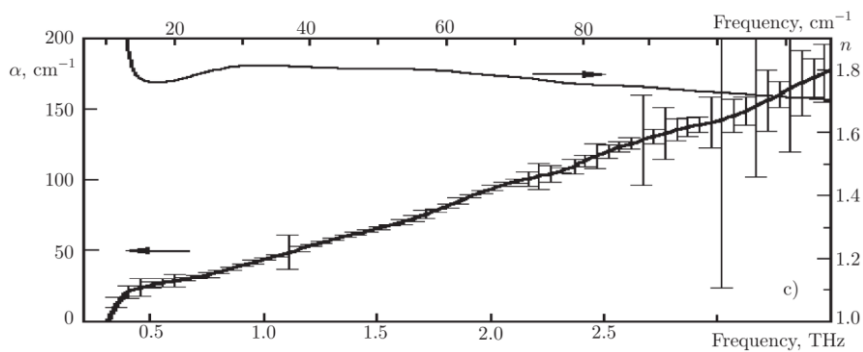

Фиг. 2. Спектр коэффициента поглощения *α* молекулы ДНК [13] (нижняя кривая, левая ордината).

*(б)* образование активных форм кислорода (АФК) в гидратных группах, окружающих молекулу ДНК (на пару азотистых оснований приходится ≈ 22 гидратных группы [14]); это следует из результатов Пельтека и др. [15-18], согласно которым воздействие ТГц излучения на бактерию *Escherichia coli* приводило к появлению в её клетках перекиси водорода, что активировало экспрессию гена, чувствительного к окислительному стрессу. Согласно Offei-Danso *et al.* [19], в воде возможны коллективные возбуждения в виде резких, «скачкообразных» (*angular jumps*), разворотов молекул воды, приводящих к разрыву водородных связей между молекулами. По-видимому, облучение молекул воды ТГц волнами способствует возникновению таких резких разворотов: ТГц волна своим электрическим полем могла бы действовать на дипольный момент молекулы воды, резко разворачивать её, разрывать водородные связи и, вероятно, генерировать АФК;

*(в)* химическое взаимодействие разорванных пар азотистых оснований и образовавшихся АФК, в результате чего в ядре клетки появляются вещества, угнетающие её жизнь.

Согласно гипотезе, представленной в работе [20] (см. *Appendix A* в ней; а также п.2 ниже), ЗНЧ размером ≤ 8 *нм* – источники спонтанного ТГц излучения. Значит, при внесении их в живую клетку в последней следует ожидать проявления следов эффектов, перечисленных выше.

Например, ЗНЧ размером 1,4 *нм*, попав в большую бороздку (*major groove*) молекулы ДНК, благодаря предполагаемой спонтанной эмиссии ТГц фотонов способны вызвать сильный биологический эффект. Этим можно воспользоваться для терапии раковой опухоли. Для проверки этого способа терапии проще воспользоваться случаем, когда опухоль видна невооружённым глазом (например, в случае асцитной карциномы Эрлиха) и поэтому доступна для инъекции ЗНЧ непосредственно в опухоль. Такое введение наночастиц упрощает процедуру их доставки к раковым клеткам (не требуются конъюгаты антител с ЗНЧ), и, поскольку способ не предполагает гипертермии раковых клеток, не требуется и лазер ближнего инфракрасного диапазона. В работе [21] ЗНЧ со средним диаметром 13 *нм* вводили непосредственно в интерстиций опухоли Эрлиха, выращенной у мыши породы *Balb* путём подкожной инъекции клеток асцитной карциномы Эрлиха. Мы можем воспользоваться этим методом доставки ЗНЧ в опухоль Эрлиха – путём инъекции ЗНЧ размером 1,4 *нм* непосредственно в неё.

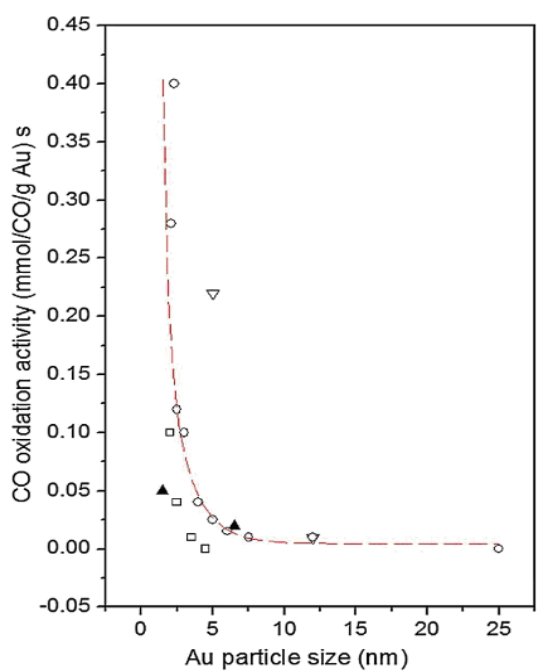

Фиг. 3. Каталитическая активность ЗНЧ (для реакции окисления монооксида углерода) в зависимости от их размера [22].

Нанокластеры золота $Au_{55}$, имеющие размер 1,4 *нм*, особенно цитотоксичны: их размер соответствует ширине 2,2 *нм* большой бороздки в молекуле ДНК [23]. Поэтому нанокластер $Au_{55}$ хорошо входит в большую бороздку и может задерживаться в ней на достаточно продолжительное время, облучая молекулу и гидратные группы вокруг неё ТГц фотонами. В работе [24] показано, что из ЗНЧ ряда размеров (0,8 *нм*; 1,2 *нм*; 1,4 *нм*; 1,8 *нм*) наночастицы именно размера 1,4 *нм* оказались наиболее цитотоксичными (на кривой каталитической активности ЗНЧ, Фиг. 3, этот размер соответствует участку с очень высокой каталитической активностью).

Косвенно в пользу гипотезы [20] о спонтанной эмиссии ТГц фотонов наночастицами золота и токсичности последних, применимой для задач онкологии, говорят результаты исследования Tsoli *et al.* [25], в котором в экспериментах *in vitro* показана токсичность ЗНЧ размером 1,4 *нм* по отношению к клеткам одиннадцати видов рака человека.

Вероятно, угнетение жизнедеятельности клеток вызывает изменение их *морфологии*, что наблюдалось в работах [26-28] при введении в организм животных ЗНЧ размерами 1-3 *нм* и 50 *нм*. Однако изменение *морфологии* клеток наблюдалось и в работе [29] из-за нарушения клеточного деления после облучения клеток ТГц фотонами. Это наводит на мысль об общей причине морфологических изменений, а именно: эмиссии ТГц излучения наночастицами золота.

Размерная зависимость цитотоксичности ЗНЧ коррелирует с размерной зависимостью их каталитической активности (Фиг. 3): ЗНЧ размером ~1-5 *нм* токсичны, в то время, как ЗНЧ размером ~6-20 *нм* не токсичны [30]. Это аргумент в пользу предположения об общей физической природе каталитической активности и цитотоксичности ЗНЧ – спонтанной эмиссии ТГц фотонов наночастицами золота меньше ~8 *нм*.

В связи с графиком на Фиг. 3 отметим, что ЗНЧ, выбранные для экспериментов *in vivo* [26-28], имели размеры 1-3 *нм*, 15 *нм* и 50 *нм*, резко отличающиеся по каталитической активности, что, вероятно, и объясняет полученные результаты. Наночастицы 1-3 *нм* с высокой каталитической активностью вызвали морфологические изменения в органах лабораторных животных; наночастицы 15 *нм*, согласно Фиг. 3, не проявляющие никакой каталитической активности, не вызвали морфологических изменений; а наночастицы 50 *нм* вызвали морфологические изменения в органах животных, вероятно, из-за локальных электрических полей контактной разности потенциалов (КРП) (см. п.2.2 ниже).

## 2. Действующие начала наночастиц золота
## 2.1. Спонтанная эмиссия ТГц фотонов

Спонтанная эмиссия ТГц фотонов золотой наночастицей диаметром меньше ≈8 *нм* является проявлением размерного эффекта и квантования уровней энергии вибрационных мод (продольных фононов) в условиях конфайнмента. Рассмотрим волну сжатия (продольный фонон с импульсом $q^*_{vm}$), распространяющуюся вдоль диаметра $D$ ЗНЧ (Фиг. 4(*а*)). Продольные фононы с энергией $E_{vm}$ и импульсом $q^*_{vm}$ постоянно поглощаются фермиевскими электронами (предполагается, что последние достаточно корректно описываются в рамках модели свободных электронов) с импульсом $p_F = 1,27 \cdot 10^{-19}$ *г·см/с* и энергией $E_F = 5,53$ *эВ* [31]. Возбуждённые (в результате поглощения фононов) электроны движутся вдоль направления, характеризуемого импульсом *s*, модуль которого равен $s = [2m \cdot (E_F + E_{vm})]^{1/2}$, под углом γ к импульсу фонона, γ = arccos $[(2m \cdot E_{vm} + q^{*2}_{vm})/(2s \cdot q^*_{vm})]$. Для доминирующего в золоте продольного фонона (с энергией $E_{vm} \approx 17,4$ *мэВ* и величиной импульса $q^*_{vm} \approx 1,13 \cdot 10^{-19}$ *г·см/с*) угол γ ≈ 63,6°.

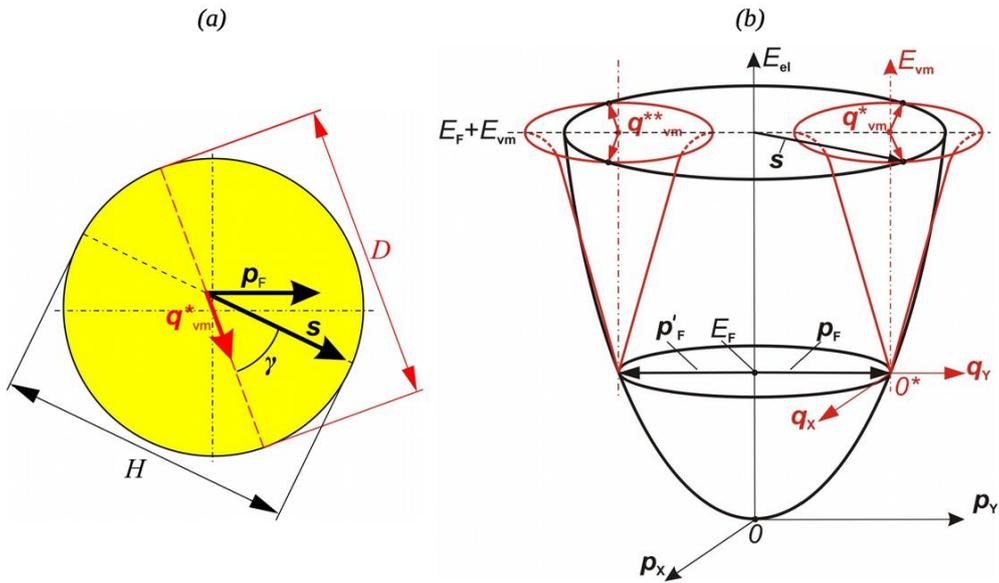

Фиг. 4. Поглощение вибрационной моды (продольного фонона) с импульсом $q^*_{vm}$ фермиевским электроном с импульсом $p_F$ в ЗНЧ. *(a)* Возбуждённый электрон с импульсом $s$ движется вдоль хорды $H$. *(b)* Дисперсионные поверхности электронов (параболоид) и продольных фононов (конусовидные поверхности с раструбами красного цвета). Из работы [20].

Возбуждённые электроны, благодаря кулоновскому взаимодействию с положительными ионами золота, могли бы индуцировать вторичные продольные фононы с импульсами, коллинеарными импульсу $s$ ; это был бы «естественный» способ релаксации для возбуждённых электронов. Однако в ЗНЧ меньше ≈8 *нм* этот механизм невозможен, и релаксация происходит путём эмиссии фотона.

Первая причина, способствующая эмиссии, – это разница в шагах квантования импульсов и энергий у продольных фононов, распространяющихся вдоль диаметра наночастицы $D$ и вдоль хорды $H$ (по которой движется возбуждённый электрон с импульсом $s$). Шаги квантования импульсов для фононов, распространяющихся вдоль диаметра и хорды, равны, соответственно, $h/D$ и $h/H$ (здесь $h$ – постоянная Планка).

Так как $D > H$, то шаги по импульсам **различны**; следовательно, шаги по энергии для этих двух направлений распространения фононов тоже **не совпадают**. Из-за несовпадения квантования уровней энергии переход энергии от возбуждённого электрона к вторичному фонону невозможен (от первичного фонона фермиевский электрон получил энергию, квантованную с определённым шагом, и он не может генерировать вторичный фонон, у которого энергия была бы квантована с другим шагом). Дисперсионная кривая фононов для направления вдоль вектора $s$, вообще говоря, не совпадает с дисперсионной кривой фононов для направления вдоль диаметра наночастицы, но это принципиально не ослабляет сделанного вывода, если не укрепляет его.

Другая причина, препятствующая возбуждению вторичного фонона и, следовательно, благоприятствующая излучению фотона, заключается в следующем. С уменьшением длины хорды $H$ зазор между энергетическими уровнями продольных фононов для направления вдоль вектора $s$ увеличивается и постепенно становится настолько большим, что превышает ширину $FWHM_L$ энергетического распределения продольных фононов в ЗНЧ (на Фиг. 5 показано пороговое состояние этой ситуации; подробности см. ниже). При этом для направления вдоль диаметра $D$ в пределах $FWHM_L$ остаётся хотя бы один уровень продольного фонона (так как $H < D$, шаги по импульсам и энергии для хорды больше шагов по импульсам и энергии для диаметра, и в пределах $FWHM_L$ найдётся первичный фонон). В результате, энергетический уровень возбуждённого электрона «повисает» между уровнями продольных фононов для направления вдоль $s$: в пределах ширины $FWHM_L$ для этого направления не существует ни одного уровня фонона, которому он мог бы передать энергию. Следовательно, электрон релаксирует при рассеянии на границе ЗНЧ, испуская фотон (электрон не может покинуть ЗНЧ, так как его энергия меньше работы выхода электрона из золота, равной ≈4,8 *эВ*). Возбудить вторичный фонон, бегущий по поверхности ЗНЧ, не удастся: электрон был возбуждён первичным фононом, квантованным по импульсам с шагом *h/D*, а у продольных фононов на поверхности ЗНЧ шаг по импульсам в π раз меньше: *h/πD*.

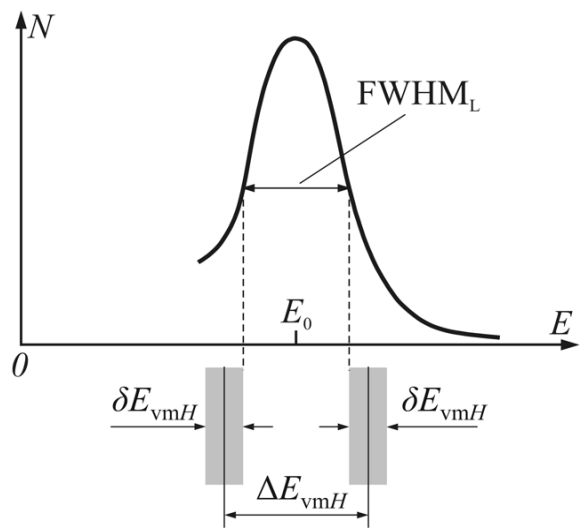

Фиг. 5. Пороговое состояние, при котором шаг по энергии *ΔE*$_{vmH}$ вибрационных мод, распространяющихся вдоль хорды *H* параллельно импульсу ***s***, превышает величину FWHM$_L$ энергетического распределения продольных фононов в ЗНЧ. Из работы [20].

Для направления «вдоль хорды *H*» шаги по импульсу и энергии вибрационных мод равны, соответственно, *h/H* и *ΔE*$_{vmH}$ ≈ $v^*_L$·(*h/H*), где $v^*_L$ – скорость звука для диапазона энергий продольных фононов в пределах FWHM$_L$. Величина $v^*_L$ меньше номинальной скорости звука в золоте, $v_L = 3{,}23·10^5$ *см/с*, из-за искривления «книзу» дисперсионной кривой продольных фононов вблизи границы зоны Бриллюэна, в каковой области и лежат энергии фононов, соответствующие полосе FWHM$_L$.

*Таблица 1.* Корреляция каталитической активности ЗНЧ и её диаметра *D*. Из работы [20].

| *D* (*нм*) | *ΔE*$_D$ (*мэВ*) | *δE*$_{vmH}$ (*мэВ*) | FWHM$_L$/*ΔE*$_D$ | Комментарий к Фиг. 3 |
|---|---|---|---|---|
| 1,2 | 3,45 | 0,55 | 0,81 | Наибольшая каталитическая активность |
| 1,5 | 2,76 | 0,44 | 1,01 | Снижение каталитической активности при увеличении диаметра *D* |
| 2,5 | 1,655 | 0,26 | 1,69 | |
| 5,0 | 0,83 | 0,13 | 3,37 | |
| 8,0 | 0,52 | 0,08 | 5,38 | |
| 8,5 | 0,49 | 0,08 | 5,71 | Отсутствие каталитической активности |
| 10,0 | 0,41 | 0,065 | 6,83 | |

Несовпадение уровней энергии отчасти смягчается соотношением неопределённостей Гейзенберга. В качестве критерия «полного несовпадения» энергетических уровней фононов, которые потенциально могли бы быть возбуждены, мы выбираем ситуацию, при которой два соседних энергетических уровня раздвигаются на энергию, превышающую ширину FWHM$_L$ с учётом «размытия» уровней энергии вследствие соотношения неопределённостей. Такая пороговая ситуация изображена на Фиг. 5, где *δE*$_{vmH}$ оценивается с использованием соотношения неопределённостей Гейзенберга для импульса фонона и его координаты, *δE*$_{vmH}$ ≥ $v^*_L$·*h*/(2π*H*). Неравенство, определяющее соответствующую пороговую длину хорды (такую, что при меньших её значениях начинается спонтанная эмиссия фотонов), выглядит так:

$$\Delta E_{\text{vm}H} > \text{FWHM}_L + (\delta E_{\text{vm}H}/2) + (\delta E_{\text{vm}H}/2) \geq \text{FWHM}_L + v^*_L \cdot h/(2\pi H). \qquad (1)$$

Подставляя величину $\Delta E_{\text{vm}H} \approx v^*_L \cdot (h/H)$ в левую часть неравенства (1), получим:

$$v^*_L \cdot (h/H) > \text{FWHM}_L + v^*_L \cdot (h/2\pi H),$$

откуда  $H < [1 - (1/2\pi)] \cdot v^*_L \cdot (h/\text{FWHM}_L).$   (2)

При выполнении этого неравенства возбуждённые электроны не смогут релаксировать с генерацией вторичных фононов, а будут испускать фотоны. Подставляя численные значения величин ($v^*_L \sim 10^5$ *см/с*; $\text{FWHM}_L \approx 2{,}8$ *мэВ*), получим для длины хорды, обеспечивающей высокую интенсивность спонтанной эмиссии фотонов: $H < 1{,}2$ *нм*.

Фермиевский электрон с импульсом $p_F$ может поглощать фонон с импульсом $q^*_{\text{vm}}$ (его длина волны больше, чем 0,408 *нм*, но меньше, чем $H_0 = 1{,}2$ *нм*) в любой точке диаметра $D$. Длина хорды минимальна, когда поглощение первичного фонона электроном происходит на поверхности наношара. В этом случае диаметр $D$ и хорда $H_0$ образуют две стороны треугольника, вписанного в окружность (с гипотенузой – диаметром), следовательно, треугольник прямоугольный, поэтому $D = H_0/\cos \gamma$. Полагая здесь $H_0 = 1{,}2$ *нм*, $\gamma \approx 63{,}6°$, получим значение диаметра ЗНЧ, начиная с которого, при уменьшении диаметра начинается резкий рост эмиссии фотонов: $D = 2{,}7$ *нм*. Это согласуется с ходом каталитической активности ЗНЧ: на Фиг. 3 при диаметрах $D < 2{,}7$ *нм* наблюдается резкий рост каталитической активности – по нашему мнению, из-за того, что при этом $H < H_0$, и энергетические уровни вибрационных мод раздвигаются на увеличивающийся зазор $\Delta E_{\text{vm}H}$, выходя за пороговое состояние, показанное на Фиг. 5 (см. также комментарий в Таблице 1).

Отметим, что энергия доминирующего в золоте продольного фонона равна $E_{\text{vm}} \approx 17{,}4$ *мэВ*; эта энергия, поглощённая фермиевским электроном и эмитированная возбуждённым электроном в виде фотона, соответствует частоте терагерцевого диапазона: 4,2 *ТГц*.

Именно спонтанной эмиссией ТГц фотонов мы объясняем природу каталитической активности, наблюдаемую у ЗНЧ (см. работы [32-34]) размером $\leq 8$ *нм* (см. Таблицу 1). Мы считаем, что эмиссия ТГц фотонов способствует катализу химических реакций на поверхности ЗНЧ. Поскольку у многих молекул вибрационные и вращательные энергии лежат в ТГц диапазоне, ТГц поля могут влиять на внутримолекулярные связи.

Эмиссия фотонов будет сопровождаться охлаждением ЗНЧ, так как последние не теплоизолированы. Прекращение излучения фотонов при $D \approx 8$ *нм* могло бы быть и проявлением диссипации энергии возбужденного электрона за счет его многократного рассеяния на других электронах, если бы размер ЗНЧ становился сравнимым со средней длиной свободного пробега электронов $l_{\text{mfp}}$ в ЗНЧ. Судя по экспериментальным данным каталитической активности ЗНЧ в зависимости от их диаметра [32-34], в ЗНЧ величина $l_{\text{mfp}} \geq 8$ *нм*.

Чем больше кратность укладки шага по энергии продольных фононов $\Delta E_D$ в пределах ширины $\text{FWHM}_L$, то есть чем больше отношение ($\text{FWHM}_L/\Delta E_D$), тем выше вероятность совпадения уровня энергии, «размытого» на $\delta E_{\text{vm}H}$, с уровнем энергии первичного фонона, что приводит к релаксации электрона с генерацией вторичного фонона (с энергией, равной энергии первичного фонона). Тем ниже интенсивность спонтанной эмиссии фотонов и ниже каталитическая активность ЗНЧ, и, соответственно, ниже степень охлаждения ЗНЧ.

Результаты работ [32-34] предполагают, что в ЗНЧ, в диапазоне диаметров $D < 8$ *нм*, не выполняется неравенство $l_{\text{mfp}} < D$ (иначе возбуждённый электрон рассеялся бы на других электронах, и эмиссия была бы невозможна). В работах [35, 36] показано, что в наночастицах серебра (химического аналога золота) размером меньше 10 *нм* сохраняется кристаллическая решётка, а её искажение происходит лишь вблизи поверхности наночастиц. Для этого случая $l_{\text{mfp}} = D$, и $D < l_{\text{bulk}}$, где $l_{\text{bulk}}$ – длина свободного пробега электронов в массивном золоте (*bulk*

*gold*); величина $l_{bulk}$ довольно велика: $l_{bulk}$ = 37,7 *нм* [37]. Всё это предполагает, что в ЗНЧ, исследованных в работах [32-34], рассеяние электронов происходит лишь на границах ЗНЧ (электрон пересекает ЗНЧ от границы до границы без рассеяния в её объёме), то есть выполняются условия для эмиссии фотонов.

Но это также подтверждает предположение, что в ЗНЧ прекращение эмиссии с ростом размера ЗНЧ связано именно с попаданием уровней вибрационных мод для направления вдоль хорды – в полосу шириной $FWHM_L$.

В связи с нашей гипотезой о спонтанной эмиссии ТГц фотонов наночастицами золота (и их охлаждении) при размерах меньше 8 *нм*, отметим, что, когда некоторые из нас изучали размерный эффект при нагреве ЗНЧ методом радиочастотной гипертермии (на частоте ~13,56 *МГц*) [38], мы не нашли данных об успешном использовании ЗНЧ с размерами меньше 5 *нм*. Это косвенное подтверждение нашей гипотезы: эффективно нагревать ЗНЧ с размерами менее 5 *нм* невозможно потому, что, будучи нагреваемы радиочастотным излучением, они одновременно охлаждались из-за спонтанной эмиссии ТГц фотонов, уносивших энергию из ЗНЧ. Это коррелирует с данными, что катализ начинается при размерах ЗНЧ ≤8 *нм*. Согласно нашим модельным оценкам [38], в ЗНЧ с размерами < 5 *нм*, всё ещё может существовать механизм нагрева, связанный с продольными фононами, распространяющимися по поверхности ЗНЧ, но его роль снижается с уменьшением диаметра ЗНЧ. По-видимому, он преодолевается охлаждением из-за спонтанной эмиссии ТГц фотонов, интенсивность которых увеличивается с уменьшением диаметра ЗНЧ, и сопровождается ростом каталитической активности ЗНЧ.

### 2.1.1. Признак в пользу гипотезы о спонтанной эмиссии ТГц фотонов

Экспериментальные данные о плотности состояний фононов для массивного образца золота и для ЗНЧ размером 5 *нм* отличаются (Фиг. 6): наблюдается «дефицит» в пике плотности состояний продольных фононов у ЗНЧ в области энергий вокруг 17,4 *мэВ* (частот 4,2 *ТГц*), см. Фиг. 6 *слева*, случай (b), и Фиг. 6 *справа*. Вероятно, этот «дефицит» отражает уменьшение

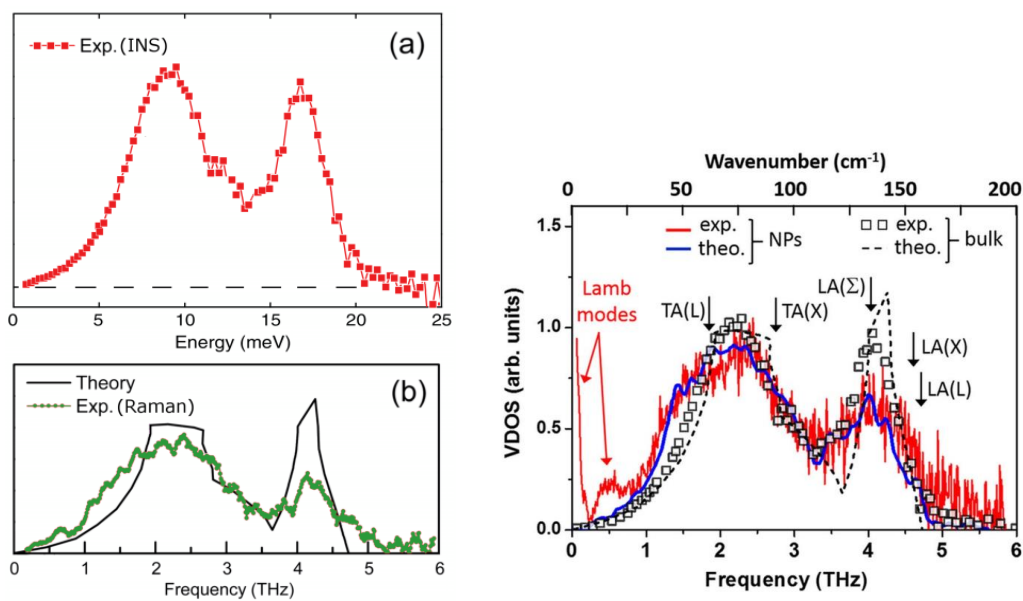

Фиг. 6. Плотность состояний фононов в золоте: *слева*: извлечённая из: *(a)* экспериментов по неупругому рассеянию нейтронов в толстой фольге [39], *(b)* плазмонного резонанса при рамановском рассеянии в нанокристаллах (зелёные точки) [40] и восстановленная из силовых констант, подогнанных к данным о неупругом рассеянии нейтронов в массивных монокристаллах [41] (непрерывная чёрная линия); *справа*: Fig. 5 из статьи [42]: плотность состояний фононов (VDOS) в массивном золоте (квадратики – эксперимент, пунктирная линия – теоретический расчёт) и у ЗНЧ (синяя линия – моделирование, красная линия – эксперимент).

плотности состояний продольных фононов вследствие квантования их энергий в ЗНЧ, а также из-за **поглощения части фононов фермиевскими электронами**. (Квантование энергий имеет место и в массивном образце, но у ЗНЧ шаг квантования намного больше, из-за чего в наночастицах многие значения энергии «выпадают» и не реализуются).

**2.1.2. Оценка плотности мощности ТГц излучения, эмитированного наночастицей золота**

Для электрона, поглотившего продольный фонон, бежавший вдоль диаметра ЗНЧ, время нахождения в возбуждённом состоянии $\Delta t$ связано с неопределённостью в энергии возбуждённого состояния $\Delta E$ соотношением Гейзенберга:

$$\Delta E \cdot \Delta t \geq \hbar. \qquad (3)$$

Неопределённость $\Delta E$ связана с неопределённостью в импульсе $\Delta p$ возбуждённого электрона по формуле: $\Delta E \approx v_F \cdot \Delta p$, где $v_F$ – скорость фермиевских электронов (см. *Appendix B* и Fig. 15 в статье [20]).

Неопределённость $\Delta p$ найдём из соотношения Гейзенберга для неопределённости импульса возбуждённого электрона $\Delta p$ и его положения $\Delta x$ на хорде длиной $H$:

$$\Delta p \cdot \Delta x \geq \hbar.$$

Поскольку $\Delta x \approx H$, то $\Delta p \geq \hbar/H$, и $\Delta E \geq v_F \cdot \hbar/H$.

Из неравенства (3) для «времени жизни возбуждённого состояния» $\Delta t$ получим: $\Delta t \geq \hbar/\Delta E$. Подставив сюда неравенство $\Delta E \geq v_F \cdot \hbar/H$, получим: $\Delta t \sim H/v_F$.

Будем считать, что «время жизни возбуждённого состояния» $\Delta t$ определяется суммой времени возбуждения электрона $\tau_1$, среднего времени пробега $<t>$ возбуждённого электрона по хорде до рассеяния на границе ЗНЧ, и времени эмиссии фотона электроном $\tau_2$ при рассеянии на границе ЗНЧ:

$$\Delta t \approx \tau_1 + <t> + \tau_2. \qquad (4)$$

Для простоты предположим, что $\tau_1 = \tau_2$, а $<t> = H/2v_F$ (здесь, в соответствие с представлениями квантовой механики, мы положили, что возбуждённый электрон как бы «размазан» вдоль хорды $H$, и поэтому время его пробега до рассеяния на границе наночастицы может лежать в интервале от 0 до $H/v_F$; поэтому его средняя величина составит $H/2v_F$). Тогда, с учётом ранее полученного соотношения $\Delta t \sim H/v_F$, выражение (4) примет вид: $H/v_F \sim 2\tau_2 + H/2v_F$. Отсюда для времени эмиссии фотона $\tau_2$ электроном получим:

$$\tau_2 \sim H/4v_F.$$

Мощность, излучаемая при эмиссии одного фотона с энергией $h\nu$, равна $P \approx h\nu/\tau_2 \approx 4h\nu \cdot v_F/H$. А соответствующая плотность мощности равна: $I \approx P/S$, где $S$ – площадь *волновой поверхности* у эмитированной электроном электромагнитной волны с длиной волны $\lambda$. На фронте волны, на расстоянии одной длины волны, вернее, $(D/2) + \lambda$, от поверхности ЗНЧ диаметром $D$ величина $S = 4\pi \cdot [(D/2) + \lambda]^2$. Для доминирующего в золоте продольного фонона с энергией 17,4 *мэВ* длина эмитированной волны равна $\lambda \approx 7,1 \cdot 10^{-3}$ *см*. Тогда для плотности мощности при $H_0 = 1,2$ *нм* получаем оценку:

$$I \approx P/S \approx 4h\nu \cdot v_F/H_0 \cdot 4\pi \cdot [(D/2) + \lambda]^2 \approx h\nu \cdot v_F/H_0 \cdot \pi \cdot \lambda^2 = hc \cdot v_F/\pi \cdot H_0 \cdot \lambda^3 = 21{,}3 \text{ мВт/см}^2.$$

Эта величина – для случая нахождения ЗНЧ в вакууме; в водной же среде клетки длина электромагнитной волны будет меньше, и, соответственно, плотность мощности – выше. Таким образом, даже грубая оценка плотности мощности ТГц излучения, обеспечиваемого одним возбуждённым электроном в ЗНЧ, даёт величину в пределах значений плотностей мощности, использованных в ряде экспериментов с облучением живых клеток ТГц фотонами (см. обзоры [43-48]), в которых наблюдался биологический эффект.

Как соотнести полученную оценку для плотности мощности с существующими нормами безопасности для ТГц излучения [49, 50]? В настоящее время не существует исчерпывающих норм безопасности, учитывающих как тепловое, так и нетепловое действие ТГц волн и регулирующих применение излучения в диапазоне от 0,1 до 10,0 *ТГц*. Воздействие излучения с частотами <0,3 *ТГц* регулируется нормами безопасности [49], а с частотами >0,3 *ТГц* – нормами [50]. Для частот <0,3 *ТГц* максимальная плотность мощности для населения ограничивается величиной 10 *Вт/м²*, а для частот >0,3 *ТГц* при длительности воздействия >10 *с* она на два порядка больше: 1 *кВт/м²*.

В работе [43] подчёркивается, что такое расхождение двух норм безопасности связано с разными правилами выбора порога воздействия. Первый стандарт определяет пределы безопасности, исходя из повышения внутренней температуры тела на 1°C (с 37°C до 38°C). А второй стандарт основан на температурном пороге повреждения кожи в 45°C. При этом оба стандарта не учитывают нетепловые эффекты ТГц волн, которые мы рассматриваем.

Непосредственное сравнение затруднительно и потому, что нормы относятся к человеческому организму в целом или отдельному органу (глазам или коже) и зависят от площади и времени облучения, а мы оцениваем воздействие ТГц излучения ЗНЧ на клетку. Например, предел плотности мощности для частот >0,3 *ТГц* относится к площади облучаемой кожи 0,01 *м²* (например, к площади кожи 10 *см* × 10 *см*). Очевидно, для объектов облучения существенно меньше (например, для живой клетки организма) этот предел намного меньше. Поэтому воздействие ТГц излучения наночастиц на клеточные объекты предпочтительнее сравнивать с результатами соответствующих экспериментов на клетках организмов.

### 2.1.3. Наночастица золота как источник импульсного ТГц излучения

Интересующие нас ЗНЧ размером меньше 2,7 *нм*, по сути, являются источниками *импульсного* ТГц излучения. Как мы уже оценили, такая ЗНЧ способна обеспечить плотность мощности в импульсе ~ 21 *мВт/см²*. А какова частота повторения этих импульсов? Каково время между актом эмиссии ТГц фотона и моментом следующего возбуждения фермиевского электрона за счёт поглощения продольного фонона?

Из-за недостатка данных для более или менее надёжных оценок возможностей ЗНЧ как импульсного источника ТГц излучения, прибегнем к прогнозу (*speculations*) путём сравнения.

В идеале следовало бы выполнять оценку, сравнивая: (1) импульсный источник ТГц излучения с известными частотой повторения импульсов и плотностью мощности ТГц излучения в импульсе и (2) ЗНЧ, которая обеспечивает такую же плотность мощности ТГц излучения в импульсе; при этом, воздействуя на один и тот же вид живых клеток одинаковое время, импульсный источник и ЗНЧ вызывают в клетках одинаковый биологический эффект.

За неимением такой возможности, выполним грубые сравнения на имеющемся примере – импульсном источнике ТГц излучения с плотностью мощности не больше, чем 21 *мВт/см²*. Известен импульсный источник ТГц излучения с частотой 1,7 *ТГц* и плотностью мощности 2,4 *мВт/см²* из статьи [51], имеющий частоту повторения импульсов 1 *кГц* и вызывающий биоэффект в виде деметилирования молекулы ДНК в клетках рака крови.

ЗНЧ размером ≤ 2,7 *нм* цитотоксичны, следовательно, при обеспечиваемой ими оценочной плотности мощности ~21 *мВт/см²* биоэффект применения этих ЗНЧ достижим при частоте повторения импульсов эмиссии ТГц фотонов, по крайней мере, ~1 *кГц*, характерной для импульсного источника из работы [51]. То есть, биоэффект мог бы иметь место при времени между актом эмиссии ТГц фотона и моментом следующего возбуждения фермиевского электрона $\tau^* \sim 10^{-3}$ *с*.

Соотнесём величину $\tau^* \sim 10^{-3}$ *с* со средним временем между актами рассеяния электрона в массивном золоте: $\tau_0 \approx l_{\text{bulk}}/v_F = 37,7$ *нм*$/1,4 \cdot 10^8$ *см/с* $\approx 2,7 \cdot 10^{-14}$ *с*; оно характеризует рассеяние электрона на обоих видах фононов (величины $l_{\text{bulk}}$ и $v_F$ – из [37] и [31], соответственно). Нас же интересует среднее время $\tau^L$ между актами рассеяния электрона на продольных фононах. Очевидно, $\tau^L > 2{,}7 \cdot 10^{-14}$ *с*, то есть в массивном золоте $l^L_{\text{mfp}} > l_{\text{bulk}}$, где $l^L_{\text{mfp}}$ – средняя длина свободного пробега электрона, обусловленная рассеянием электрона на продольных фононах.

Трудно найти физический механизм в ЗНЧ, благодаря которому время между актами рассеяния электрона на продольных фононах радикально – на несколько порядков величины – увеличилось бы с $\tau^L > 2{,}7 \cdot 10^{-14}$ *с* до $\tau^* \sim 10^{-3}$ *с*. Один из физических механизмов увеличения времени $\tau^L$ в ЗНЧ можно связать с тем, что частота рассеяния электрона на продольных фононах в ЗНЧ меньше, чем в массивном кристалле – потому, что число продольных фононов в ЗНЧ меньше из-за снижения плотности состояний.

Действительно, по сравнению с массивным кристаллом, в ЗНЧ вдоль её хорд и диаметра не могут распространяться фононы с длиной волны, превышающей длину хорд и диаметра, и в энергетическом распределении фононов состояния таких фононов отсутствуют. «Прореживанию» энергетического распределения фононов способствует и квантование импульсов и энергий фононов, распространяющихся вдоль хорд и диаметра.

Если принять высказанное предположение, что по сравнению с массивным кристаллом, в ЗНЧ время $\tau^L$ увеличивается, и допустить, что соответственно увеличивается и $l^L_{\text{mfp}}$ (и превышает длину свободного пробега $l^L_{\text{mfp}}$ в массивном золоте), и вследствие этого $l^L_{\text{mfp}} > D$, то это означало бы, что возбуждённый электрон в ЗНЧ в интервале между актами неупругого рассеяния многажды упруго (то есть без передачи энергии) рассеивается на границе ЗНЧ – со средней периодичностью $t_{\text{ср}} \sim D/2v_F$, где $D/2$ – среднее расстояние, пролетаемое электроном в ЗНЧ диаметром $D$ между двумя актами упругого рассеяния; $t_{\text{ср}} \ll \tau^L$.

В этом случае частота повторения импульсов эмиссии ТГц фотонов, определяемая увеличенными значениями $l^L_{\text{mfp}}$ и $\tau^L$, снизилась бы. Если бы она всё же оказалась выше ~1 *кГц*, это означало бы, что ЗНЧ имеют преимущество перед импульсными источниками ТГц излучения – они могут вызывать биоэффекты, будучи более цитотоксичными по отношению к раковым клеткам. Это был бы ещё один аргумент в пользу предположения, что ЗНЧ размером $\leq 2{,}7$ *нм* – эффективные импульсные источники ТГц излучения.

**2.2. Локальные электрические поля контактной разности потенциалов**

Наночастицы золота, вероятно, могут вызывать биологический эффект в живых клетках не только за счёт спонтанной эмиссии ТГц фотонов, но и из-за высоких значений напряжённости локальных электрических полей КРП. Действительно, у наночастиц размером > 1,4 *нм* (с числом атомов больше 70) работа выхода электрона с ростом размера ЗНЧ стабилизируется и монотонно приближается к работе выхода массивного кристалла [52] (Фиг. 7).

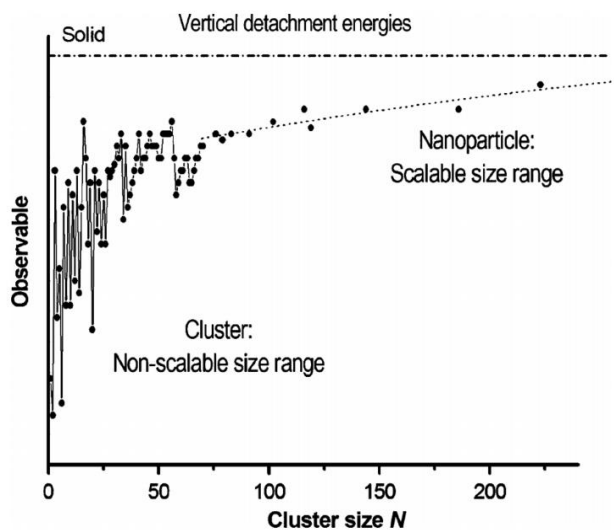

Фиг. 7. Работа выхода электрона из кластеров золота с разным числом атомов [52].

Для быстрого перевода диаметра $D_{MS}$ наночастиц золота в число атомов золота $N_{Au}$ и обратно удобно пользоваться кривой на Фиг. 8 (см. верхнюю абсциссу), заимствованной из работы [53]. Например, ЗНЧ диаметром 2,7 *нм* содержит ≈600 атомов, и, согласно Фиг. 7, у неё можно ожидать величин работ выхода, типичных для массивного кристалла золота.

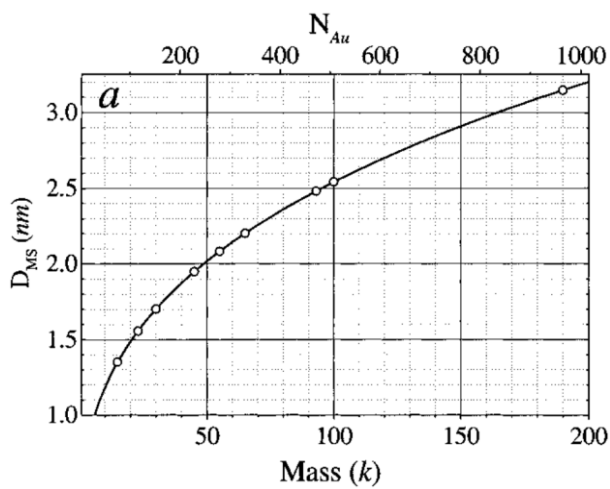

Фиг. 8. Взаимосвязь между диаметром $D_{MS}$ наночастицы золота и числом атомов золота $N_{Au}$ в ней (см. верхнюю абсциссу). Из статьи [53].

Ответ на вопрос, с какого размера ЗНЧ является кристаллом, можно найти в работах [54-57] и на Фиг. 9, где ясно различимы область крупных ЗНЧ (область обратных размеров < 0,6 *нм*$^{-1}$; диаметры $D > 1,7$ *нм*) и область малых ЗНЧ (область обратных размеров > 0,6 *нм*$^{-1}$; $D < 1,7$ *нм*), в которой температура плавления ЗНЧ почти не зависит от их размера. Считается [54-57], что в области $D > 1,7$ *нм* ЗНЧ являются кристаллами. Это не противоречит данным работ [35, 36], в которых показано, что в наночастицах серебра, химического аналога золота, размером меньше 10 *нм* кристаллическая решётка сохраняется, а её искажение происходит лишь вблизи поверхности наночастиц. Из работ [54-57] также следует, что при $D < 1,7$ *нм* ЗНЧ – не кристаллические, и представляют собой так называемое «наностекло» [55]. К этой области относятся и нанокластеры $Au_{55}$ размером 1,4 *нм*.

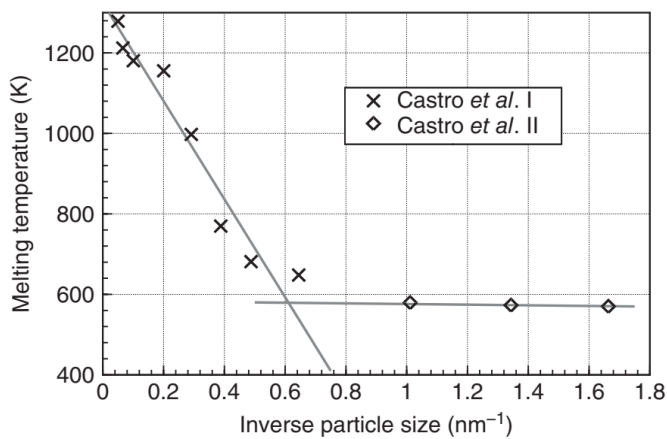

Фиг. 9. Температура плавления ЗНЧ как функция их обратного размера (из работы [54]).

В ЗНЧ с $D \geq 1,7$ *нм* можно ожидать различий в работах выхода электрона с разных граней нанокластера золота, а следовательно, и присутствия электрических полей КРП – над рёбрами и вершинами ЗНЧ. Действительно, предположив, что в ЗНЧ с $D \geq 1,7$ *нм* сохраняется разность работ выхода с соседних граней приблизительно такая же, как в массивных кристаллах золота (~ 0,1 *эВ*), см. Таблицу 2, при характерном расстоянии в 1 *нм* между двумя точками на соседних гранях нанокластера, над рёбрами, разделяющими соседние грани, напряжённость электрического поля составит величину $E_{КРП}$ ~ 0,1 В/1 нм = $10^6$ В/см.

*Таблица 2.* Величины работ выхода электрона из граней монокристалла золота с малыми индексами Миллера [58]

| Грань | Работа выхода *(эВ)* | Неопределённость в работе выхода *(эВ)* |
|---|---|---|
| Au(100) | 5,22 | 0,31 |
| Au(111) | 5,33 | 0,06 |
| Au(110) | 5,16 | 0,22 |

Такого же порядка напряжённости электрического поля могут иметь место и над вершинами нанокластера золота. Для сравнения: напряжённость, соответствующая электрической прочности воздуха при нормальных условиях, равна ≈ 3·$10^4$ *В/см*, то есть гораздо меньше напряжённости электрического поля над рёбрами и вершинами наночастицы золота. Электропрочность воздуха определяется пробоем в результате диссоциации молекулы газа в электрическом поле. По-видимому, и гетерогенный катализ на ЗНЧ связан с диссоциацией молекул, участвующих в химической реакции на наночастице, которая ещё и спонтанно излучает ТГц фотоны, способные воздействовать на вращательные и колебательные состояния молекул (их энергии лежат в ТГц диапазоне).

Таким образом, в живой клетке следы воздействия локальных электрических полей КРП, могут проявиться у ЗНЧ с размерами ≥ 1,7 *нм*.

На Фиг. 10 показана огранка наномонокристалла золота размером ≈ 4 *нм*. Подобная огранка наблюдается и у нанокластеров размером ~ 2 *нм* [59].

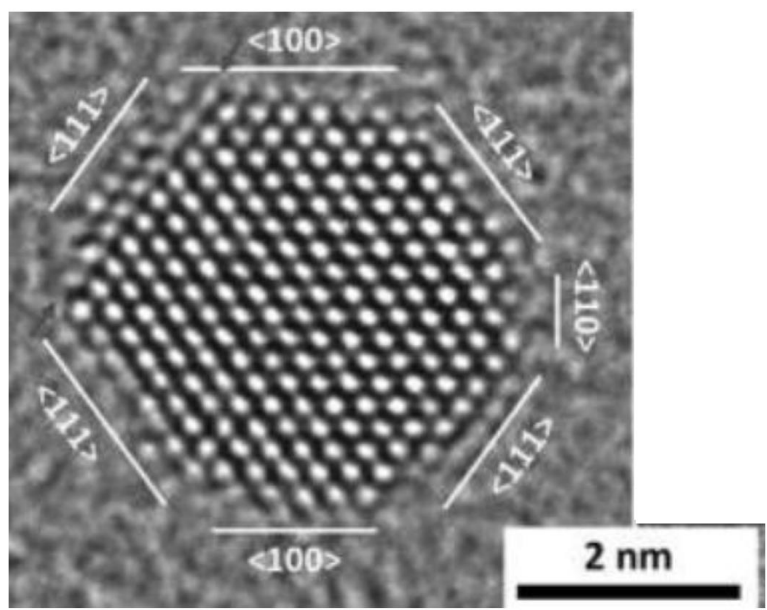

Фиг. 10. Электронно-микроскопическое изображение наномонокристалла золота. Указаны кристаллографические обозначения граней. Из работы [59].

Таким образом, ЗНЧ размером меньше 1,7 *нм* – не кристаллическая, и ЗНЧ размером 1,4 *нм*, способная войти в большую бороздку молекулы ДНК, наиболее эффективная для задач онкологии – тоже не кристаллическая; их цитотоксичность обусловлена эмиссией ТГц фотонов. ЗНЧ размером меньше 2,7 *нм* – очень интенсивно излучает ТГц фотоны. Поэтому ЗНЧ с диаметрами 1,7 *нм* < D < 2,7 *нм* – могут как испускать ТГц фотоны, так и иметь локальные электрические поля КРП. У ЗНЧ > 2,7 *нм* с ростом диаметра снижается интенсивность эмиссии ТГц фотонов, но растёт длина рёбер, на которых имеют место локальные электрические поля КРП.

О роли напряжённости электрического поля, создаваемого ТГц излучением в живой клетке, и вызывающего в ней биологический эффект:

*(а)* Fröhlich [60, 61] считал, что биологический эффект достижим при напряжённости электрического поля $E_{Fr} \geq 3\cdot10^4$ *В/см*;

*(б)* деметилирование молекулы ДНК в раковых клетках крови происходит при напряжённости электрического поля $2{,}8\cdot10^5$ *В/см* [51];

*(в)* в экспериментах Ситникова и др. [62-65] напряжённости электрического поля составляли ~ $10^6$ *В/см*, и в них наблюдался биологический эффект (можно предположить, что в воде цитоплазмы и кариоплазмы импульсы ТГц излучения могли разрывать водородные связи и, возможно, генерировать АФК, вызывающие окислительный стресс);

*(г)* при напряжённости электрического поля $2{,}5\cdot10^7$ *В/см* ТГц излучение ускоряет размотку цепочек в двухцепочечной молекуле ДНК [66] – очевидно, из-за разрыва водородных связей между азотистыми основаниями.

В связи с мнением Fröhlich'а [60, 61], что биоэффект достижим при $E_{Fr} \geq 3\cdot10^4$ *В/см*, и учитывая, что напряжённости электрического поля $E_{КРП}$ у ЗНЧ (даже крупной, ≥10 *нм*, не эмитирующей ТГц фотоны) – того же порядка, что и в экспериментах Ситникова и др. [62-65] (~$10^6$ *В/см*), можно ожидать, что, так как $E_{КРП} > E_{Fr}$, наночастицы в живой клетке могли бы вызывать биоэффекты в виде окислительного стресса (например, путём генерации АФК из воды цитоплазмы и кариоплазмы – за счёт диссоциации молекул воды, по аналогии с диссоциацией молекул газа в воздухе). И действительно, есть работы [23, 24, 67-72], в которых в живые клетки вносили ЗНЧ и отмечали окислительный стресс в клетках.

Отметим работу [73], подтверждающую наши предположения, что: *(i)* благодаря ЗНЧ размером 1,4 *нм*, проникшей в ядро клетки, в последней образуются АФК (их признаком является *окислительный стресс*, отмеченный в этой работе); *(ii)* АФК в «пузырях» молекулы ДНК взаимодействуют с разорванными азотистыми основаниями, образуя угнетающие клетку соединения (в работе отмечается *некроз* клеток); *(iii)* у ЗНЧ размером 1,4 *нм* цитотоксичность намного выше, чем у ЗНЧ размером 15 *нм* (что согласуется с графиком на Фиг. 3).

Эффективность воздействия статических электрических полей КРП на клетку связана с размером ЗНЧ, типом клетки, вероятностью проникновения ЗНЧ в клетку и её ядро, а также, возможно, и с другими факторами.

В работе [74] исследовали поглощение живыми клетками ЗНЧ размерами от 2,4 до 89 *нм*. Обнаружилось, что, во-первых, поглощение зависит от проявления на поверхности ЗНЧ свойств способствующего проникновению в клетку пептида и, во-вторых, глубина проникновения ЗНЧ внутрь клетки зависит от её диаметра. Наименьшие ЗНЧ, размером 2,4 *нм*, локализовались в ядре клетки; ЗНЧ с промежуточными размерами 5,5 и 8,2 *нм* проникали в цитоплазму, не доходя до ядра, а часть ЗНЧ оставалась в мембране. ЗНЧ размером 16 *нм* и больше – не проникали в клетки и располагались на их периферии. Выводы из этой работы в целом согласуются с результатами исследований [75-77]: в работе [75] отмечено проникновение ЗНЧ размерами 2 и 6 *нм* как в ядра раковых клеток, так и в цитоплазму, а у ЗНЧ размером 15 *нм* – лишь в цитоплазму; в работе [76] ЗНЧ размерами 2 и 6 *нм* были обнаружены в ядрах клеток, а более крупные ЗНЧ размерами 10 и 16 *нм* – только в цитоплазме; авторы статьи [77] тоже отмечали способность ЗНЧ размером 2 *нм* проникать в ядра клеток.

Экспериментально наблюдаемая цитотоксичность ЗНЧ коррелирует с поглощательными свойствами живых клеток в отношении ЗНЧ. Действительно, ЗНЧ размером 1–3 *нм*, способные проникать в ядра клеток, наиболее цитотоксичны; ЗНЧ размером 15 *нм* либо застревают в мембране, либо не проникают в неё (и каталитически не активны) и не вызывают морфологических изменений (хотя в других экспериментах ЗНЧ размером 15 *нм* могут вызывать окислительный стресс – это говорит о том, что свойства ЗНЧ зависят не только от их размера, но и от других факторов: типа клеток, вещества лиганда на поверхности ЗНЧ и т.п.). Аналогично, ЗНЧ размером 50 *нм* не эмитируют ТГц фотоны, и, казалось бы, они не должны проявлять цитотоксичности. Однако существуют публикации, в которых показано, что такого размера ЗНЧ могут вызывать окислительный стресс – см., например, работу [70], в которой использовались ЗНЧ размерами 30, 50 и 90 *нм*. Авторы показали, что цитотоксичность этих ЗНЧ вызвана окислительным стрессом, определяемым количеством генерируемых ими АФК, и из испытанных в этой работе ЗНЧ максимальная генерация АФК наблюдалась у наночастиц размером 50 *нм*. Это коррелирует с результатами работы [78], в которой исследовалась проникающая способность в клетки у ЗНЧ размерами 14, 30, 50, 74 и 100 *нм*, и обнаружилось, что именно наночастицы размером 50 *нм* имеют наибольшую проникающую способность в клетки.

Хотя можно предположить, что наблюдавшаяся в работе [70] максимальная генерация АФК у ЗНЧ размером 50 *нм* обусловлена их максимальной проникающей способностью в клетки [78], мы полагаем, что первопричиной генерации АФК этими наночастицами являются локальные электрические поля КРП на рёбрах и вершинах наночастиц. Именно ими можно объяснить морфологические изменения в органах лабораторных животных при использовании ЗНЧ размером 50 *нм*, отмеченные в работах [26-28].

Как уже отмечено, для исследования воздействия ЗНЧ на опухоль карциномы Эрлиха мы планируем доставлять ЗНЧ непосредственно в опухоль путём инъекции наночастиц размером 1,4 *нм*. Для уверенного выявления роли спонтанного ТГц излучения наночастиц в угнетении раковых клеток (и проверки размерного эффекта в нашей гипотезе), целесообразно провести также эксперименты с инъекцией ЗНЧ размерами 15 и 50 *нм*. Исследование следует выполнить, сравнивая образцы биоткани одного типа, взятых у: *(а)* мыши, не привитой карциномой Эрлиха; *(б)* мыши, привитой карциномой Эрлиха; *(в)* мыши, привитой карциномой Эрлиха с последующим введением в опухоль ЗНЧ трёх размеров: 1,4; 15; и 50 *нм*.

В обзоре [10] авторы перечислили три главных причины цито- и генотоксичности ЗНЧ, которые совпали с приведёнными нами во Введении тремя эффектами в живых клетках, облучаемых ТГц фотонами. Но авторы не связали причины токсичности ЗНЧ со спонтанной эмиссией ими ТГц фотонов.

### 2.2.1. Признак в пользу гипотезы о локальных электрических полях КРП

В дополнение к рассмотренным данным о ЗНЧ размером 50 *нм* заслуживает внимания недавно опубликованная статья [79], анализ результатов которой может прояснить не только детали *интернализации* ЗНЧ в живую клетку в зависимости от размера ЗНЧ, но также выявить роль явлений, способствующих проникновению ЗНЧ в клетку и окислительному стрессу, а главное – увидеть признак, подтверждающий нашу гипотезу о локальных статических электрических полях КРП на рёбрах и вершинах наночастиц золота.

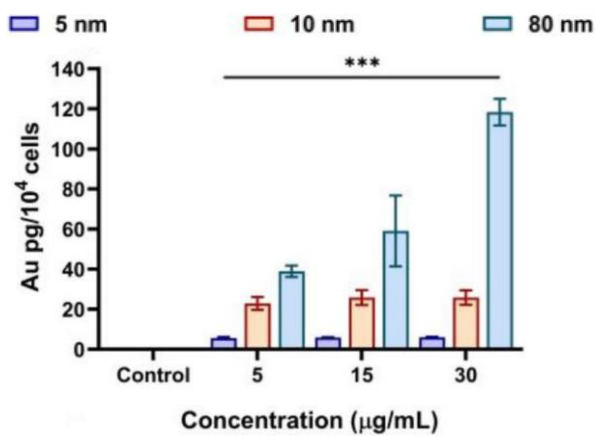

Фиг. 11. Проникающая способность ЗНЧ трёх размеров (5, 10 и 80 *нм*) в клетки А549 после 24 часов инкубации (по абсциссе отложена концентрация ЗНЧ в инкубационном растворе в единицах *мкг/мл*; по ординате – масса ЗНЧ, проникших в клетки в единицах *пикограмм* на $10^4$ клеток). Из статьи [79].

Из гистограмм Фиг. 11 видно, что проникающая способность ЗНЧ в живые клетки тем больше, чем больше размер наночастицы, и она растёт с концентрацией ЗНЧ в инкубационном растворе.

По-видимому, ЗНЧ размером < 10 *нм* проникают в клетку, проходя сквозь липидный бислой мембраны, лишь «раздвигая» фосфолипиды (*direct membrane penetration*); ЗНЧ размером > 10 *нм* проникают в клетку благодаря *эндоцитозу*, не повреждая мембрану клетки; а ЗНЧ размером ~ 10 *нм* проникают в клетку, дестабилизируя целостность липидного бислоя, что приводит к частичной гибели клеток. Это объясняет, почему на гистограмме Фиг. 12 ЗНЧ размером 10 *нм* по проценту жизнеспособных клеток занимают промежуточное положение между ЗНЧ размерами 5 и 80 *нм*. Процент жизнеспособных клеток минимален для наночастиц 5 *нм* – из-за их высокой цитотоксичности вследствие эмиссии ТГц фотонов, генерации АФК и проникновения в ядра клеток, согласно работам [75, 76]. Процент жизнеспособных клеток максимален у клеток, поглотивших наночастицы размером 80 *нм*, благодаря следующему:

*(а)* согласно Фиг. 11, у наиболее цитотоксичных наночастиц размерами 5 и 10 *нм* проникающая способность в клетку ниже, чем у наночастиц 80 *нм* (и поэтому в клетке ЗНЧ размером 80 *нм* больше, чем ЗНЧ с размерами 5 и 10 *нм*);

*(б)* наночастицы 10 *нм* и 80 *нм* своими локальными электрическими полями КРП на рёбрах и вершинах генерируют АФК в клетке (у ЗНЧ 80 *нм* общая длина рёбер и, соответственно, область локальной генерации АФК больше, чем у ЗНЧ размером 10 *нм*). Казалось бы, выживаемость клеток, содержащих ЗНЧ 80 *нм* должна быть ниже, чем у клеток с ЗНЧ размером 10 *нм*. Но включается механизм гомеостаза клетки, отвечающий за поддержание АФК в норме, и клетке с ЗНЧ 80 *нм* удаётся выживать *ценой морфологических изменений* в ней (по-видимому, аналогичные *морфологические изменения* наблюдались в работах [26-28] при применении ЗНЧ размером 50 *нм*). У клеток же, поглотивших ЗНЧ размером 10 *нм*, очевидно, вероятность погибнуть выше – потому, что при проникновении таких ЗНЧ в клетку нарушается целостность клеточной мембраны.

В результате гомеостаза, Фиг. 13 отражает динамические значения генерируемых АФК в клетках. Очевидно, это сказалось и на проценте жизнеспособных клеток с ЗНЧ 80 *нм* (Фиг. 12) – без работы механизма гомеостаза процент выживших клеток, поглотивших ЗНЧ 80 *нм*, на Фиг. 12 был бы меньше.

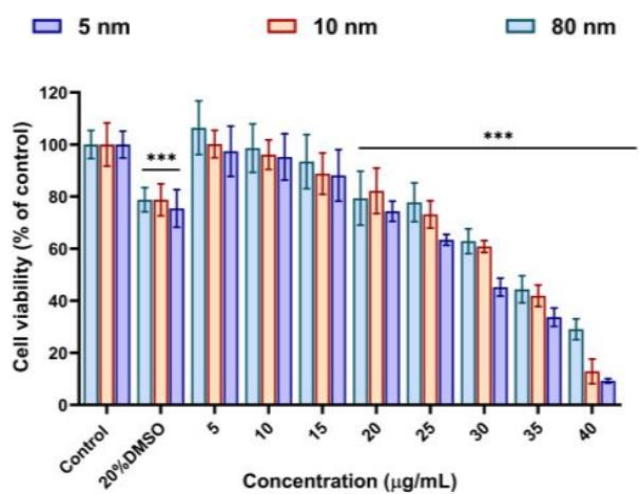

Фиг. 12. Процент жизнеспособных клеток A549 после их инкубации в течение 24 часов в растворе, содержащем ЗНЧ трёх размеров: 5, 10 и 80 *нм*. По оси абсцисс отложена концентрация ЗНЧ в инкубационном растворе в единицах *мкг/мл*; по оси ординат – процент жизнеспособных клеток A549 относительно контрольного раствора. Из статьи [79].

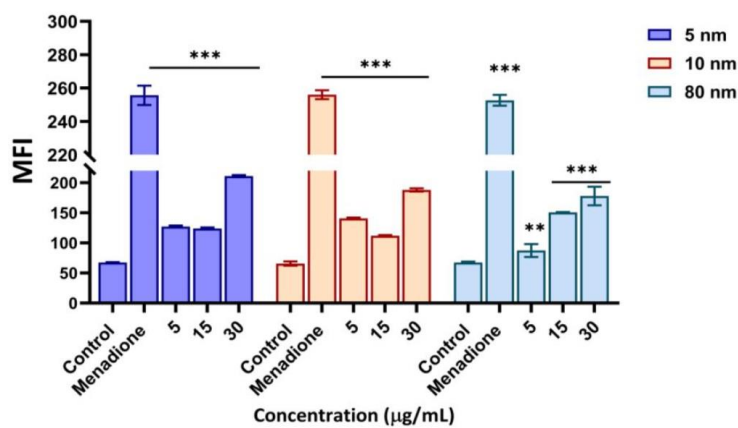

Фиг. 13. Генерация АФК в клетках A549 при воздействии на них наночастицами золота размерами 5, 10 и 80 *нм* (по оси абсцисс отложена концентрация наночастиц в инкубационном растворе в единицах *мкг/мл*; по оси ординат – средняя интенсивность флуоресценции, отражающая интенсивность генерации АФК в клетках). Из статьи [79].

Авторы работы [79] измеряли уровень генерации АФК, регистрируя среднюю интенсивность флуоресценции в живых клетках, то есть клетки могли реагировать на генерируемые АФК – у них включался механизм гомеостаза клетки, отвечающий за поддержание АФК в норме. Поэтому измеренные результаты отражают реакцию клетки, и без ресурсов клетки по гомеостазу концентрация АФК в клетках, поглотивших ЗНЧ размером 80 *нм*, была бы выше, чем следует из гистограмм на Фиг. 13, а процент выживших клеток на Фиг. 12 – ниже.

Для авторов статьи [79] было неожиданно, что даже с учётом статистической погрешности, начиная с концентрации 15 *мкг/мл*, генерация АФК в клетках у ЗНЧ размером 80 *нм* – выше, чем у ЗНЧ размером 10 *нм*. Мы видим в этом признак в пользу гипотезы о локальных электрических полях КРП на рёбрах и вершинах наночастиц золота – уровень генерации АФК соответствовал большей общей длине рёбер у ЗНЧ 80 *нм* по сравнению с общей длиной рёбер у ЗНЧ размером 10 *нм*.

Как отмечалось, и Фиг. 13 подтверждает это, производство АФК наночастицей 5 *нм* велико при всех концентрациях наночастиц в инкубационном растворе, это согласуется с цитотоксичностью этих ЗНЧ вследствие спонтанной эмиссии ТГц фотонов и их способностью проникать в ядра клеток.

**3. Обзор экспериментальных данных о сперматотоксичности наночастиц золота**

Мужская репродуктивная система очень чувствительна к облучению электромагнитным излучением, которое в сперматозоидах вызывает генерацию АФК и повреждает молекулы ДНК (см., например, [80, 81]). Поэтому следы эмиссии ТГц излучения наночастицами золота могут обнаружиться в опытах со сперматозоидами. В связи с этим интересны эксперименты Захидова и др. [82-85], в которых использовались ЗНЧ размерами 2-3 *нм* – фактически они выступили в роли источников ТГц излучения, а сперматозоиды – в качестве естественных сенсоров этого излучения. Результаты опытов можно расценить как аргумент в пользу нашей гипотезы о спонтанной эмиссии ТГц фотонов наночастицами золота размером $\leq 8$ *нм*.

Захидов с коллегами исследовали воздействие ЗНЧ размерами 2-3 *нм* на сперматозоиды животных. В этих работах, выполненных *in vitro* путём инкубации сперматозоидов в растворе, содержащем ЗНЧ, была обнаружена сперматотоксичность наночастиц золота. Выводы из экспериментов [82, 83] на сперматозоидах самцов мыши: ЗНЧ диаметром ~2,5 *нм* в концентрациях $1,0 \cdot 10^{15}$ и $0,5 \cdot 10^{15}$ *частиц/мл* сперматотоксичны, причём механизм воздействия ЗНЧ связан с их взаимодействием с молекулами двухспиральной ДНК; в этих опытах количество зрелых гамет с полностью деконденсированными ядрами достигало 100% против 44% в контроле.

Изучение влияния ЗНЧ размерами 2-3 *нм* на хроматин сперматозоидов мыши [84] показало: после инкубации сперматозоидов в растворе с ЗНЧ, в сперматозоидах происходит частичная или полная деконденсация ядерного хроматина. Морфологически (по степени распухания ядер и распаковки хроматина) гаметы, инкубированные в присутствии ЗНЧ, значительно отличались от контрольных. После инкубации в растворе с ЗНЧ доля сперматозоидов, достигших максимальной степени деконденсации ядер, была ≈80% по сравнению с ≈10% в контроле.

В работе [85] изучали реакцию эякулированных сперматозоидов быка на ЗНЧ методом деконденсации ядерного хроматина *in vitro*. После обработки образцов спермы в растворе, содержащем ЗНЧ со средним диаметром ~3 *нм* и концентрацией $1 \cdot 10^{15}$ *частиц/мл*, способность ядер сперматозоидов к деконденсации резко изменилась. В сперматозоидах, обработанных наночастицами золота, было отмечено появление большого количества гамет с деструктированными и почти полностью разрушенными ядрами.

В работе Wiwanitkit *et al.* [86], также выполненной *in vitro* путём инкубации сперматозоидов человека в растворе с ЗНЧ, тоже была обнаружена сперматотоксичность наночастиц золота. Воздействуя на сперматозоиды наночастицами золота размером ≈9 *нм*, авторы обнаружили, что в сперме, смешанной с раствором ЗНЧ, 25% сперматозоидов не были подвижны. Наблюдалось проникновение ЗНЧ в головку и хвостики сперматозоидов, а также фрагментация сперматозоидов. Результаты можно объяснить окислительным стрессом из-за повышенного уровня АФК, генерируемых благодаря локальным электрическим полям КРП на рёбрах и вершинах наночастиц, проникших в клетки.

Теперь – о непосредственном воздействии ТГц излучения на сперматозоиды. В единственном найденном нами исследовании эффекта ТГц излучения на сперматозоиды человека, [87], выполненном так же *in vitro* (на частотах 0,1-3 *ТГц* с плотностью мощности 60 *мкВт/см$^2$*), обнаруженный авторами биологический эффект ТГц облучения – увеличение подвижности сперматозоидов (эффект длился более 60 минут после прекращения облучения), при этом не нарушалась целостность молекул ДНК. Вероятно, этот результат объясняется низкой плотностью мощности применённого ТГц излучения (она не превышала допустимую величину 1 *мВт/см$^2$* согласно [49]).

**4. Об осуществимости поставленной задачи**
Предпосылки к возможности экспериментального обнаружения проявлений предложенных здесь физических явлений в клетках живых организмов связаны с возможностью изготовления заведомо чистых наночастиц золота. Существует технология производства чистых ЗНЧ, исключающая фоновое действие на клетки остаточных токсичных веществ-примесей в наночастицах (как это может быть при использовании ЗНЧ, полученных методами коллоидной химии). Это технология фемтосекундной лазерной абляции золота, последовательно разрабатываемая Кабашиным [88-92]. Другое достоинство этой технологии: ЗНЧ, полученные этим способом, имеют узкий разброс размеров.

**5. Подготовительные сравнительные эксперименты**
Как сказано выше, для проверки рассмотренной идеи целесообразно выбрать эксперименты с карциномой Эрлиха. Мы выполнили подготовительные сравнительные эксперименты, которые позволили бы в дальнейшем обнаружить изменения в биоткани раковой опухоли – после введения в неё ЗНЧ. Ниже в качестве примера приводятся микрофотографии изменений после прививки карциномы Эрлиха лишь в одном виде биоткани (остальные микрофотографии будут опубликованы в отдельной статье).

Были использованы две белые лабораторные мыши (самцы) породы *Balb*, массой 15-20 *г*, возрастом 2 месяца, с привитой асцитной карциномой Эрлиха. Штамм введенной карциномы: EαD. Каждой мыши карцинома вводилась одной инъекцией объёмом 0,2 *мл* в количестве ≈$3 \cdot 10^6$ раковых клеток. Карцинома прививалась внутрибрюшинно 13.12.2022, гибель обеих мышей зафиксирована 29.12.2022.

Мыши были получены из вивария Лаборатории биохимических основ фармакологии и опухолевых моделей НИИ ЭДиТО при ФГБУ «НМИЦ онкологии им. Н.Н. Блохина» Минздрава России. Лабораторные мыши содержались на стандартном корме и имели свободный доступ к питьевой воде. Работы проводились в соответствии с правилами лабораторной практики, утвержденными приказом Министерства Здравоохранения и Социального развития РФ от 23 августа 2010 г. №708н «Об утверждении правил лабораторной практики», Правилами лабораторной практики в РФ – приказ МЗ РФ от 19.06.2003 № 267, Директивой 2010/63/EU Европейского Парламента и Совета Европейского Союза от 22 сентября 2010 года «О защите животных, используемых для научных целей» и Международными рекомендациями по проведению медико-биологических исследований с использованием животных [93, 94].

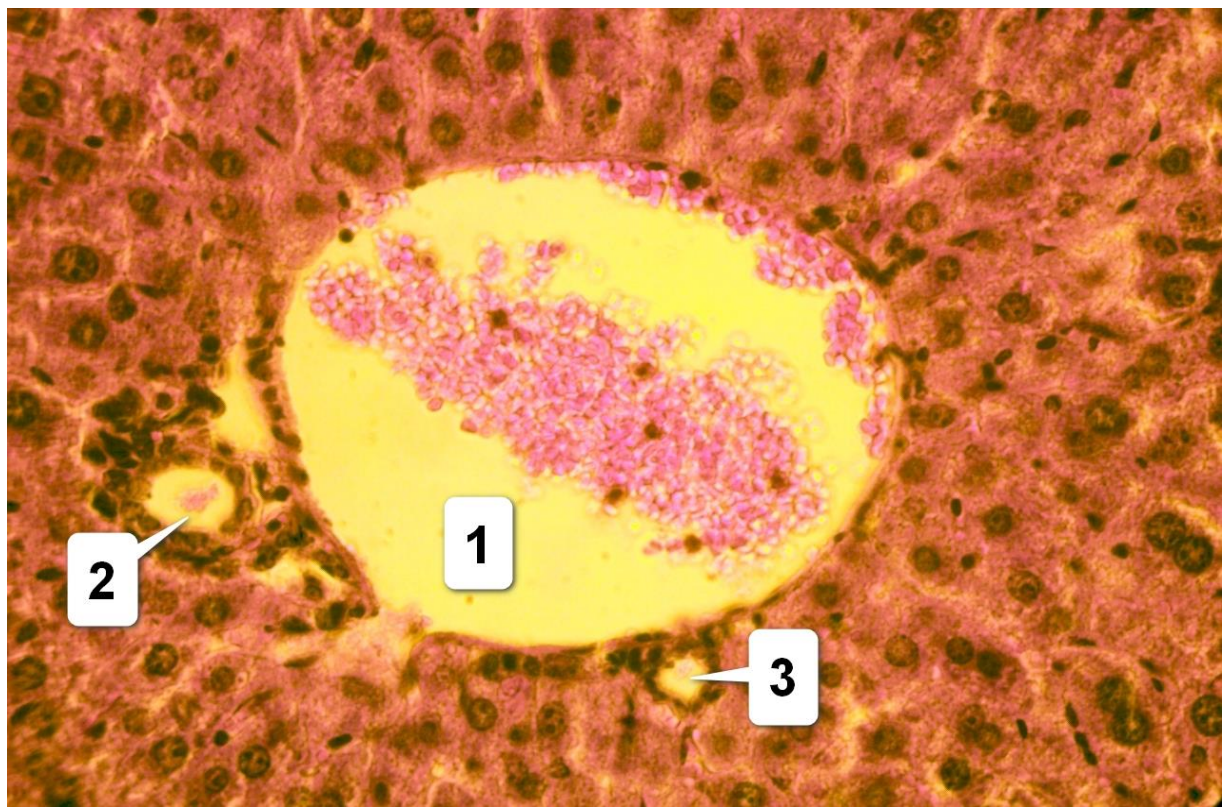

Фиг. 14. Портальная область печени белой лабораторной мыши породы *Balb* в норме. Триада печени: 1 – воротная вена, 2 – междольковая артерия, 3 – желчный проток. В воротной вене видна кровь. Хорошо дифференцируется междольковая артерия. Четко выражен желчный проток. Ширина фотографии соответствует 1280 *мкм*. Заливка в парафин.

В печени мыши с карциномой Эрлиха (Фиг. 15) наблюдаются трудно дифференцируемые гепатоциты различной формы. Сам гепатоцит увеличен в размере, как и его ядро, с изменением соотношения размера ядра к цитоплазме. В некоторых клетках ядра гепатоцитов сливаются в одно. Наряду с увеличением ядра наблюдается изменение его контура, наличие бугристости; таким образом, можно наблюдать клеточный полиморфизм. В центральной вене кровь сепарирована на форменные элементы и плазму крови. В одной центральной вене стенка с выстилающим ее однослойным плоским эпителием сохранена, в другой же центральной вене стенка распалась с образованием «частокола» из эпителия. Стенка центральной вены растянута с утратой её эластичности. Артерия увеличена в размере, её мышечная стенка расслоена, растянута. В просвете сосуда также произошла сепарация крови, но уже с образованием микротромбов. Просматривается граница, проходящая между гепатоцитами и расслоившейся артерией. Желчный проток трудноразличим, с частичной утратой однослойного цилиндрического эпителия, выстилающего его.

Исходя из этих данных и сравнительной характеристики печени здоровой мыши с печенью мыши с карциномой Эрлиха, можно сделать заключение о наличии злокачественного процесса в печени.

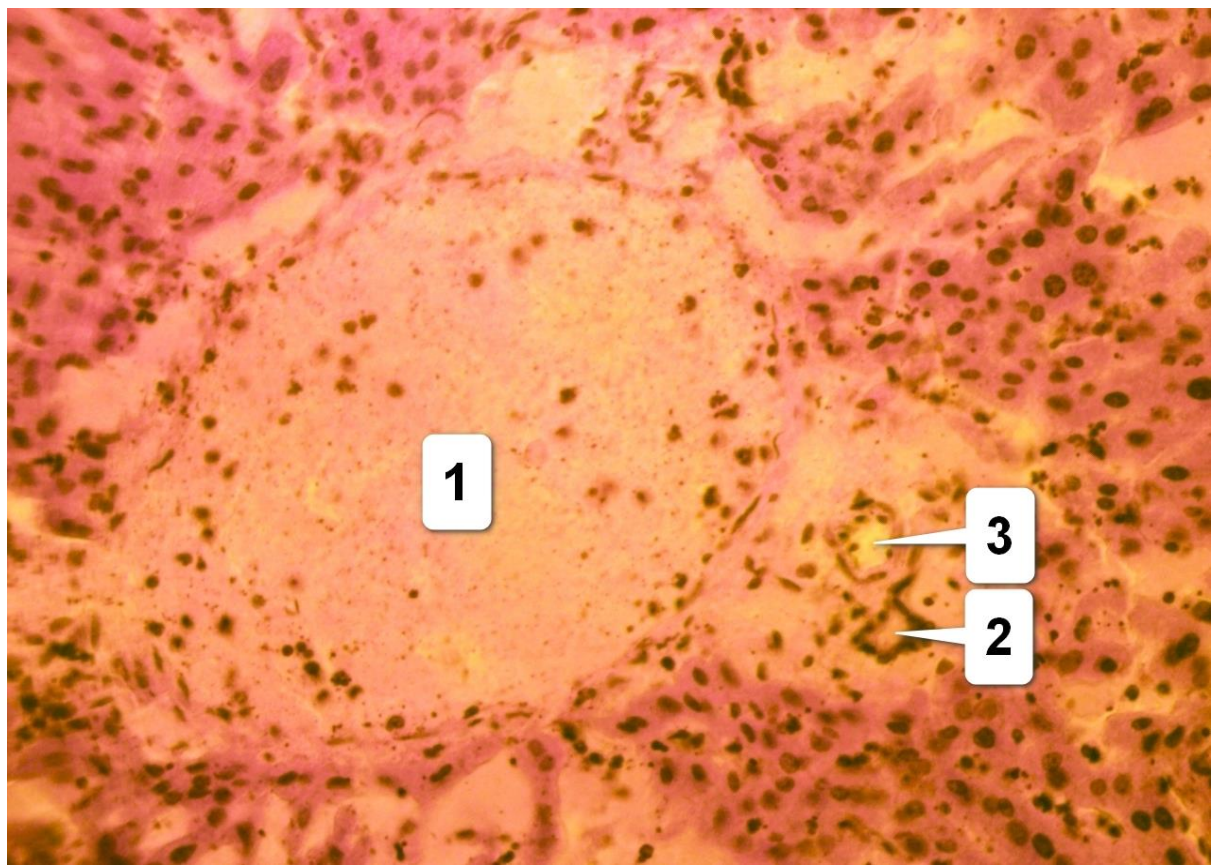

Фиг. 15. Печень белой лабораторной мыши породы *Balb* с карциномой Эрлиха. Триада печени: 1 – воротная вена, 2 – междольковая артерия, 3 – желчный проток. Ширина фотографии соответствует 1280 *мкм*. Заливка в парафин.

## 6. Заключение
В работе предложено объяснение физико-химическим механизмам токсичности наночастиц золота в живых клетках. До сих пор объяснение существующих экспериментальных данных о цито-, гено- и сперматотоксичности наночастиц золота сводят в основном к окислительному стрессу, степень которого зависит от:
– проникающей способности наночастиц в клетку, определяемой их размером, веществом-лигандом на поверхности наночастиц и типом клетки и биологической ткани;
– места в клетке, которого удаётся достигнуть наночастице в соответствие с её размером: большая бороздка молекулы ДНК (наночастицы размером 1,4 *нм*), кариоплазма ядра клетки (~2-5 *нм*), цитоплазма клетки (~7-15 *нм*) или мембрана (~20-100 *нм*).

Мы же предложили учитывать ещё два новых физических механизма, которые прежде не принимались во внимание: спонтанную эмиссию терагерцевых фотонов наночастицами золота (при размерах наночастиц ≤ 8 *нм*) и локальные электрические поля контактной разности потенциалов над рёбрами и вершинами наночастиц золота (для наночастиц размером ≥ 1,7 *нм*). На наш взгляд, учёт этих эффектов дополняет важными деталями картину взаимодействия наночастиц золота с элементами клетки.

Оценка порядка величины плотности мощности ТГц излучения, создаваемого наночастицами золота размером ≤ 2-3 *нм*, показывает, что они действительно могут быть цито-, гено- и сперматотоксичными.

В соответствие с предсказаниями нашей гипотезы, которые согласуются с экспериментальными данными работ [22, 23, 25], для поражения раковых клеток целесообразно выбирать наночастицы золота размером 1,4 *нм* как наиболее токсичные. Для проверки гипотезы, которая предсказывает размерно-зависимую токсичность наночастиц, эксперименты с внесением наночастиц золота непосредственно в опухоль карциномы Эрлиха на мышах следует проводить с наночастицами золота разного размера, а именно: 1,4 *нм*, 15 *нм* и 50 *нм*, относящимися к диапазонам размеров с отличающимися физическими свойствами.

Задача состоит в том, чтобы проследить за морфологическими изменениями ткани опухоли в зависимости от размера наночастиц, внесённой дозы наночастиц и времени, прошедшего после инъекции наночастиц в опухоль. Это было бы информативно для разработки последующих стратегий терапии с использованием наночастиц золота.

Мы рассматриваем данную работу лишь как начальный этап поиска проявлений двух действующих начал наночастиц золота в биомедицинских экспериментах с внесением наночастиц золота в живую клетку. Весьма информативными были бы эксперименты, выполненные на модельных бактериях *Escherichia coli* с биосенсорами, чувствительными к появлению активных форм кислорода в клетках, которые использовались в опытах Пельтека и др. [15-18]. Регистрация окислительного стресса в клетках бактерии *E.coli* с биосенсорами при внесении наночастиц золота в них могла бы подтвердить гипотезу о спонтанной эмиссии ТГц излучения наночастицами золота размером меньше 8 *нм*. Внедрение наночастиц золота в бактерии *E.coli*, вероятно, можно было бы выполнить путём инкубации бактерий в растворе, содержащем наночастицы золота, – аналогично тому, как это было сделано в работах [82-86].

Эксперименты с карциномой Эрлиха выбраны потому, что они позволили бы выполнить опыты *in vivo* с минимальным воздействием очень токсичных наночастиц размером 1,4 *нм* на здоровые ткани. Это позволило бы убедиться в цитотоксичности таких наночастиц не только *in vitro* (как это показано в [25]), но и *in vivo*. Следующим шагом были бы исследования наночастиц 1,4 *нм* в оболочке-лиганде, экранирующем их токсичность – такие наночастицы, доставляемые в опухоль, были бы перспективны для терапии онкозаболеваний внутренних органов.

В заключение выделим характерные области размеров наночастиц золота с краткой характеристикой вызываемого ими биологического эффекта в живых клетках (Таблица 3).

*Таблица 3.* Краткая характеристика цитотоксичности ЗНЧ в живых клетках в зависимости от размера ЗНЧ.

| Размер ЗНЧ, *нм* | Особенности биоэффекта | Цитотоксичность | Особенности ЗНЧ и физического механизма |
|---|---|---|---|
| 1,4 | Проникновение в клетку путём непосредственного пересечения мембраны; встраивание ЗНЧ в большую бороздку молекулы ДНК; формирование локальных разрывов спиралей в молекуле ДНК; окислительный стресс из-за генерации активных форм кислорода. | Максимальная | Наночастица представляет собой кластер $Au_{55}$ из 55 атомов золота. По другой классификации, ЗНЧ – «наностекло». Спонтанная эмиссия ТГц фотонов. |
| 1,7 *нм* < D < 2,7 *нм* | Проникновение в клетку путём непосредственного пересечения мембраны; проникновение в ядро клетки; окислительный | Высокая | Наночастица – кристаллическая. Повышенная интенсивность эмиссии ТГц фотонов (оценочные параметры: плотность мощности излучения ~21 *мВт/см$^2$* при частоте |
| | стресс из-за генерации активных форм кислорода. | | повторения импульсов >> 1 *кГц*). Начинают вносить вклад в цитотоксичность локальные электрические поля контактной разности потенциалов на рёбрах и вершинах ЗНЧ. |
| 3–8 | Проникновение в клетку путём непосредственного пересечения мембраны; снижение биоэффекта с увеличением размера ЗНЧ. | Монотонное снижение токсичности ЗНЧ с увеличением её размера. | Вклад в токсичность из-за локальных электрических полей контактной разности потенциалов на рёбрах и вершинах ЗНЧ ещё низок. |
| 10–15 | Проникновение в клетку с нарушением целостности липидного бислоя мембраны или за счёт эндоцитоза. | Токсичность за счёт нарушения целостности мембраны клеток. | Отсутствие эмиссии ТГц фотонов. Цитотоксичность обусловлена локальными электрическими полями контактной разности потенциалов на рёбрах и вершинах ЗНЧ. |
| 15–100 | Проникновение в клетку за счёт эндоцитоза. | Повышенная цитотоксичность при $D \approx 50$ *нм*. | |

## ЛИТЕРАТУРА